\documentclass[twocolumn,showpacs,preprintnumbers,amsmath,amssymb,floatfix,prd]{revtex4-1}
\usepackage{epsfig}
\usepackage{dcolumn}
\def \be  {\begin{equation}}
\def \ee  {\end{equation}}
\def \ba  {\begin{eqnarray}}
\def \ea  {\end{eqnarray}}
\def \baa {\begin{eqnarray*}}
\def \eaa {\end{eqnarray*}}
\def \nn {\nonumber}
\newcommand\as{\alpha_s}
\newcommand\f{\frac}
\begin{document}
\title{Fragmentation Functions at Next-to-Next-to-Leading Order Accuracy}
\begin{flushright}
LA-UR-15-28093 \\
\end{flushright}
\author{Daniele P.\ Anderle}
\email{daniele-paolo.anderle@uni-tuebingen.de}
\affiliation{Institute for Theoretical Physics, University of T\"ubingen, Auf der Morgenstelle 
14, 72076 T\"ubingen, Germany}
\author{Felix Ringer}
\email{f.ringer@lanl.gov}
\affiliation{Theoretical Division, MS B283, Los Alamos National Laboratory, Los Alamos, NM 87545, USA}
\affiliation{Institute for Theoretical Physics, University of T\"ubingen, Auf der Morgenstelle 
14, 72076 T\"ubingen, Germany}
\author{Marco Stratmann}
\email{marco.stratmann@uni-tuebingen.de}
\affiliation{Institute for Theoretical Physics, University of T\"ubingen, Auf der Morgenstelle 
14, 72076 T\"ubingen, Germany}

\begin{abstract}
We present a first analysis of parton-to-pion fragmentation functions at next-to-next-to-leading order
accuracy in QCD based on single-inclusive pion production in electron-positron annihilation.
Special emphasis is put on the technical details necessary to
perform the QCD scale evolution and cross section calculation in Mellin moment space.
We demonstrate how the description of the data and the theoretical uncertainties
are improved when next-to-next-to-leading order QCD corrections are included.
\end{abstract}


\maketitle

\section{Introduction and Motivation}
%
Within the framework of perturbative QCD (pQCD), cross sections may be written in terms of perturbatively calculable 
hard-scattering coefficient functions convoluted with appropriate sets of non-perturbative
but universal input functions constrained by data. 
The underlying theoretical foundations have been established in factorization theorems~\cite{ref:fact}.
In this work, we consider processes with identified hadrons in the final-state,
specifically, single-inclusive electron-positron annihilation (SIA) $e^+e^-\to hX$, where $h$  
denotes the detected hadron and $X$ the remaining, unidentified hadronic remnant. 
The information of how energetic quarks and gluons that are produced in SIA or other
hard-scattering processes subsequently hadronize is encoded in non-perturbative parton-to-hadron 
fragmentation functions (FFs)~\cite{ref:collins-soper}.
When considering scattering processes involving also hadrons in the initial-state, 
parton distributions functions (PDFs), the space-like counterparts of FFs,
need to be considered as well. 

A reliable quantitative description of inclusive hadron yields within pQCD crucially depends on the precise knowledge of FFs
and their uncertainties. In general, these functions are obtained from data through global QCD analyses of certain reference 
processes \cite{ref:kretzer,ref:other-ffs,ref:hkns,ref:dss,ref:dss2,ref:dssnew}.
Here, SIA data are of utmost importance, similar to the singular role played by deep-inelastic scattering (DIS) 
measurements in determinations of PDFs.
Recently, results from the {\sc Belle}~\cite{ref:belledata} and {\sc BaBar}~\cite{ref:babardata} collaborations
have complemented the existing suite of SIA data mainly from the CERN-LEP experiments taken at
a center-of-mass system (c.m.s.) energy of $\sqrt{S}=91.2\,\mathrm{GeV}$.
Thanks to the unprecedented precision of the new data sets, where the statistical uncertainties 
are mainly at the sub-percent level despite their fine binning, and the lower $\sqrt{S}$,
global QCD analyses can now utilize the energy dependence of the SIA data in the range
from about $10.5\,\mathrm{GeV}$ to $91.2\,\mathrm{GeV}$ \cite{ref:dssnew}
to extract FFs also from scaling violations, a key prediction of pQCD.

In order to match the increasing precision of the experimental data sets, the theoretical framework needs to be advanced 
as well. So far, all global fits of FFs \cite{ref:kretzer,ref:other-ffs,ref:hkns,ref:dss,ref:dss2,ref:dssnew}
were carried out at most at next-to-leading order (NLO) accuracy with still rather sizable theoretical uncertainties
due to the truncation of the perturbative series.
In this work, we present for the first time an analysis of SIA data at next-to-next-to-leading order (NNLO) in QCD,
a level already routinely accomplished in current PDF sets \cite{ref:pdfnnlo} and needed for precision CERN-LHC phenomenology. 
To reach full NNLO accuracy also for FFs, one first of all needs to include the two-loop coefficient functions
for SIA given in Ref.~\cite{ref:sia-neerven-long-as2,ref:nnlo-sia,ref:nnlo-sia+mellin1,ref:nnlo-sia-mellin2}.
In addition, the FFs exhibit a factorization scale dependence that can be calculated within pQCD
and which is governed by a set of coupled equations analogous to the Dokshitzer-Gribov-Lipatov-Altarelli-Parisi (DGLAP) 
evolution equations for PDFs. The required three-loop evolution kernels 
at ${\cal O}(\alpha_s^3)$ in the strong coupling can be found in Ref.~\cite{ref:nnlo-kernel}.

In our phenomenological study of SIA data in terms of FFs, we adopt the technical framework used in the
DSS global analyses~\cite{ref:dss,ref:dss2,ref:dssnew} which we extend to NNLO accuracy. 
As we shall discuss in some detail below, we apply efficient Mellin space techniques in order to both solve the 
evolution equations and compute the SIA cross section at NNLO. 
As it turns out, the numerical implementation of the Mellin inverse transformation, needed to compare to data,
requires special attention in case of the time-like scale evolution of FFs.
We perform global fits to SIA data at leading order (LO), NLO, and NNLO accuracy to demonstrate 
the anticipated reduction in theoretical
uncertainties inherent to the truncation of the perturbative calculation at a given fixed order in $\alpha_s$. 
We also outline how the quality
of the fit gradually improves by including higher order terms in the global analysis.
We note that first reference results for the scale evolution of FFs 
at ${\cal O}(\alpha_s^3)$ were obtained in~\cite{ref:apfelmela} with which we compare. 
For the time being, we refrain from including other sources of hadron production data used in the DSS 
global analyses at NLO accuracy \cite{ref:dss,ref:dss2,ref:dssnew},
hadron multiplicities in semi-inclusive DIS and high transverse momentum hadron production in proton-proton collisions, 
due to the lack of corresponding NNLO partonic cross sections. As a consequence, our fits will use less free parameters
than in the DSS global analyses. 

The remainder of the paper is organized as follows: in the next Section, we outline all the 
necessary technical ingredients for the extension of the pQCD framework for SIA to NNLO, specifically,
those related to the proper Mellin space implementation and the Mellin inverse transformation. 
In Sec.~\ref{sec:fit}, we briefly recall the DSS global analysis framework and discuss the
results of our fits of SIA data up to NNLO accuracy. In particular, we demonstrate the 
reduction of the scale uncertainty when increasing the perturbative order from LO and NLO to NNLO.
In addition, we compare the resulting fragmentation functions to those obtained by DSS \cite{ref:dssnew}
and Kretzer \cite{ref:kretzer}.
We summarize our main results in Sec.~\ref{sec:conclusions}, where we also discuss potential further
improvements of the presented analysis framework for fragmentation functions.

\section{Semi-inclusive $\mathbf{e^+e^-}$ annihilation up to NNLO accuracy \label{sec:nnlo}}
%
In this Section we review the necessary technical aspects to compute SIA cross sections up to NNLO 
accuracy. Special emphasis is put on the transformation from momentum to Mellin moment space and the additional 
subtleties appearing beyond NLO. 
To set the stage, we first recall in Sec.~\ref{subsec:xsec} the general structure of the SIA cross section. 
Next, we discuss some relevant features of the NNLO coefficient functions. In Sec.~\ref{subsec:evol} we review the
time-like evolution equations at NNLO and their truncated and iterated solutions, which we shall compare
numerically in Sec.~\ref{sec:fit}. Section~\ref{subsec:num} is devoted to a detailed discussion of
the numerical implementation of the Mellin space expressions and the proper choice of contour for the
Mellin inverse transformation. We will also compare to the results of the {\sc Mela} 
evolution code presented in Ref.~\cite{ref:apfelmela}. 

\subsection{Cross Section and Coefficient Functions \label{subsec:xsec}}
%
We consider the SIA process $e^+e^-\to\gamma/Z\to h X$ 
mediated by an intermediate virtual photon $\gamma$ or $Z$ boson at a c.m.s.\ energy $\sqrt{S}$, 
more specifically, hadron multiplicities defined as
\be\label{eq:TL}
\frac{1}{\sigma_{\mathrm{tot}}}
\frac{d\sigma^h}{dz} = \frac{1}{\sigma_{\mathrm{tot}}} \left[ 
\f{d\sigma^h_T}{dz}+\f{d\sigma^h_L}{dz} \right] \, .
\ee
Since we have already integrated over the scattering angle $\theta$
of the produced hadron $h$ in (\ref{eq:TL}), parity-violating interference terms 
vanish, and the cross section $d\sigma^h/dz$ can be decomposed only 
into a transverse ($T$) and a longitudinal ($L$) part, where $T, L$
refer to the $\gamma/Z$ polarizations \cite{ref:nason}.
The scaling variable $z$ is defined in terms of the four momenta $P_h$ and $q$ of the
observed hadron and $\gamma/Z$ boson, respectively, as
\be
z\equiv \frac{2P_h\cdot q}{Q^2}\,, 
\ee
where $Q^2 \equiv q^2 = S$,
and reduces to the scaled hadron energy $z=2E_h/\sqrt{S}$ in the $e^+e^-$ c.m.s.\ frame.
Experimental results for Eq.~(\ref{eq:TL}) are often given in terms of the scaled hadron
three momentum $x_p = 2p_h/\sqrt{S}$, which coincides with $z$ as long as hadron mass effects
are negligible.

Up to NNLO accuracy, i.e., ${\cal O}(\alpha_s^2)$ in the strong coupling, the total hadronic cross section $\sigma_{\mathrm{tot}}$ in Eq.~(\ref{eq:TL}) is given 
by~\cite{ref:nnlo-sia,ref:nnloint}
\ba\label{eq:nnlotot}
\sigma_{\mathrm{tot}} &=& \sigma_0 N_c\sum_q \hat{e}_q^2 
\left[ 1+ 3 C_F\,  a_s   + a_s^2 \left(-\f{3}{2} C_F^2 \right. \right. \nn \\
&&  +\, C_A C_F \left(-11\log\left(\f{Q^2}{\mu_R^2}\right) - 44 \, \zeta(3) +\f{123}{2} \right) \nn \\
&& \left. \left. +\, N_f C_F T_f\left( 4 \log\left(\f{Q^2}{\mu_R^2} \right)+16\,\zeta(3)-22 \right) \right) \right]  \, , \nn \\ 
\ea
where $\sigma_0= 4\pi\alpha^2/(3Q^2)$ is the lowest order QED cross section for $e^+e^-\to \mu^+ \mu^-$,
$\alpha$ denotes the electromagnetic fine structure constant, $\hat{e}_q$ are the electroweak quark charges, 
and $N_c=3$ is the number of colors. In addition, we have introduced the usual QCD color factors $C_A=3$,
$C_F=4/3$, and $T_f=1/2$. The sum in (\ref{eq:nnlotot}) runs over $N_f$ active massless quark flavors.
Here and throughout this paper, we use the definition $a_s=\alpha_s(\mu_R^2)/4\pi$, where
$\mu_R$ is the renormalization scale.
We refrain from reproducing the well-known expressions for the electroweak quark charges
which can be found, e.g., in Ref.~\cite{ref:nnlo-sia}.

The NNLO QCD corrections to the transverse and longitudinal cross sections
$d\sigma^h_k/dz$, $k=T,L$, in Eq.~(\ref{eq:TL}) were calculated 
in~\cite{ref:sia-neerven-long-as2,ref:nnlo-sia,ref:nnlo-sia+mellin1}. 
Adopting the same notation, they can be expressed in factorized form as
a convolution of appropriate combinations of quark and gluon fragmentation functions $D_{l=q,g}^h(z,\mu^2)$
and calculable coefficient functions $\mathbb{C}_{k,l}^{\mathrm{S,NS}}(z,Q^2/\mu^2)$:
\ba\label{eq:nnlostructure}
\f{d\sigma^h_k}{dz} = \sigma_{\mathrm{tot}}^{(0)} && \left[D_\mathrm{S}^h(z,\mu^2)\otimes \mathbb{C}_{k,q}^\mathrm{S}\left(z,\f{Q^2}{\mu^2}\right) \right. \nn \\
&& \left. + \, D_g^h\left(z,\mu^2\right)\otimes \mathbb{C}_{k,g}^\mathrm{S}\left(z,\f{Q^2}{\mu^2}\right) \right] \nn \\
&& + \,\sum_q \sigma_q^{(0)}\,D_{\mathrm{NS},q}^h(z,\mu^2) \otimes  \mathbb{C}_{k,q}^\mathrm{NS}\left(z,\f{Q^2}{\mu^2}\right)  \, , \nn\\
\ea
where, for simplicity, we have set the renormalization scale $\mu_R$ equal to
the factorization scale $\mu_F$, i.e., $\mu_R=\mu_F\equiv \mu$.
The symbol $\otimes$ denotes the standard convolution integral defined as
\be\label{eq:convolution}
f(z)\otimes g(z)\equiv \int_0^1dx\int_0^1dy \, f(x)\, g(y)\,\delta(z-xy)\, .
\ee
$\sigma_q^{(0)}$ in Eq.~(\ref{eq:nnlostructure}) is the total quark production cross section for
a given flavor $q$ at LO, ${\cal O}(\as^0$), and $\sigma_{\mathrm{tot}}^{(0)}$
is the corresponding sum over all $N_f$ active flavors. They read
$\sigma_q^{(0)} =  \sigma_0 N_c \hat{e}_q^2$ and
$\sigma_{\mathrm{tot}}^{(0)}=\sum_q\sigma_q^{(0)}$.
Factorization in Eq.~(\ref{eq:nnlostructure}) holds in general only in the presence of a hard scale,
in this case $Q$. Higher-twist corrections to Eq.~(\ref{eq:nnlostructure}),
that are suppressed by inverse powers of the hard scale, can be usually safely neglected as long as
$Q$ is large enough. We do not consider them in this study.

The non-perturbative but universal FFs $D_i^h(z,\mu^2)$ have a formal definition as
bilocal operators \cite{ref:collins-soper} and parametrize the hadronization of a
massless (anti)quark or gluon, $i=q,\bar{q},g$, into the observed hadron $h$
as a function of its fractional momentum $z$. The fragmentation 
process is assumed to be independent of any other colored particles produced in a hard scattering.
The scale dependence of the FFs is calculable in pQCD and governed by renormalization group equations
similar to those for PDFs.
The SIA cross section in Eq.~(\ref{eq:nnlostructure}) depends on the
gluon-to-hadron FF $D_g^h(z,\mu^2)$ and the
quark singlet ($\mathrm{S}$) and non-singlet ($\mathrm{NS}$) combinations that
are defined as
\be 
\label{eq:sing}
D_{\mathrm{S}}^h(z,\mu^2) =  \f{1}{N_f}\sum_q \left[ D_q^h(z,\mu^2) + D_{\bar q}^h(z,\mu^2) \right]
\ee
and
\be
\label{eq:nsing}
D_{\mathrm{NS},q}^h(z,\mu^2)  =  D_q^h(z,\mu^2) + D_{\bar q}^h(z,\mu^2) - D_{\mathrm{S}}^h(z,\mu^2)
\ee
respectively, in terms of the quark plus antiquark FFs $D_q^h(z,\mu^2) + D_{\bar q}^h(z,\mu^2)$  
for each flavor $q$.

The corresponding $i=\mathrm{S, NS}$ coefficient functions 
in Eq.~(\ref{eq:nnlostructure})
can be calculated perturbatively in pQCD as a series in $a_s$,
\be \label{eq:coeffexp}
\mathbb{C}_{k,l}^{i} = \mathbb{C}^{i,(0)}_{k,l}  + a_s\,\mathbb{C}^{i,(1)}_{k,l} + a_s^2\, \mathbb{C}^{i,(2)}_{k,l} + \ldots  \, ,
\ee
where we have suppressed the arguments $(z,Q^2/\mu^2)$ in (\ref{eq:coeffexp}).
Results are available up to ${\cal O}(a_s^2)$ \cite{ref:sia-neerven-long-as2,ref:nnlo-sia,ref:nnlo-sia+mellin1}
which is NNLO for the $\mathbb{C}_{T,l}^{\mathrm{S,NS}}$ but formally
only of NLO accuracy for the subleading longitudinal coefficient functions $\mathbb{C}_{L,l}^{\mathrm{S,NS}}$.
The latter coefficients vanish at ${\cal O}(a_s^0)$, and their perturbative series is hence shifted by
one power in the strong coupling $a_s$. The situation is completely analogous to DIS but, unlike in DIS \cite{ref:fl-dis-nnlo}, 
the ${\cal O}(a_s^3)$ NNLO contributions have not been calculated yet for SIA.
In our phenomenological studies in Sec.~\ref{sec:fit}, we will therefore resort, 
for the time being, to the approximation where the perturbative orders for $\mathbb{C}_{L,l}^{\mathrm{S,NS}}$ are counted 
as for $\mathbb{C}_{T,l}^{\mathrm{S,NS}}$, i.e., we treat the ${\cal O}(a_s^2)$ longitudinal coefficients as NNLO.
In that case, the gluon FFs does not contribute directly in SIA at LO as also $\mathbb{C}_{T,g}^{i,(0)}=0$, again, similar to DIS.
In addition, we note that up to NLO accuracy, the relation $\mathbb{C}_{k,q}^{\mathrm{S}}=\mathbb{C}_{k,q}^{\mathrm{NS}}$ holds,
which can be used to simplify Eq.~(\ref{eq:nnlostructure}) as was done, e.g., in Ref.~\cite{ref:sia-nlo}.

Numerically, in particular, when fitting a large number of data in a global QCD analysis, it is advantageous
to work in complex Mellin $N$ moment space rather than with expressions like Eq.~(\ref{eq:nnlostructure}) containing
one or several time-consuming convolution integrals.
In general, the Mellin transform $f(N)$ of a function $f(z)$ is defined by
\begin{equation}
\label{eq:mellin}
f(N)=\int_0^1dz\,z^{N-1}f(z) \, .
\end{equation}
It has the well-known property that convolutions of two functions factorize into ordinary products, i.e.,
both the transverse and longitudinal cross section $d\sigma^h_k/dz$ in Eq.~(\ref{eq:nnlostructure})
can be schematically written as products of the Mellin $N$ moments of FFs and coefficient functions,
$D^h_l(N,\mu^2) \cdot \mathbb{C}_{k,l}(N,Q^2/\mu^2)$.

The Mellin moments of the NNLO coefficient functions $\mathbb{C}_{k,l}^{i,(2)}$ in (\ref{eq:coeffexp})
were computed in both Ref.~\cite{ref:nnlo-sia+mellin1} and \cite{ref:nnlo-sia-mellin2}. 
We analytically checked the consistency of the two results, which are presented using somewhat different notations,
by independently calculating the Mellin moments from scratch starting from the 
$z$-space expressions given in Appendix C of Ref.~\cite{ref:nnlo-sia+mellin1}. 
To this end, two {\sc Mathematica} packages \cite{ref:mtpackage, ref:hplpackage} were employed.
The $z$-space results in \cite{ref:nnlo-sia+mellin1}
are given in terms of harmonic polylogarithms expressed in the notation $H_{m_1,\dots,m_w}$, $m_j=0,\pm1$ 
introduced in \cite{ref:remiddi}.
Their Mellin transform can be written in terms of harmonic sums
\begin{equation}
\label{eq:harmonicsums}
S_{a_1,\dots,a_n}(N)=\sum^{N}_{k_1=1}\!\sum^{k_1}_{k_2=1}\!\!\dots\!\!
\sum^{k_{n-1}}_{k_n=1}\frac{\text{sign}(a_1)^{k_1}}{k_1^{|a_1|}}\dots\frac{\text{sign}(a_n)^{k_n}}{k_n^{|a_n|}}\;,
\end{equation}
where the $a_k$ are positive or negative integers, and $N$ is a positive integer.
The number $n$ of $a_k$ indices indicates the so-called depth, 
whereas $w=\sum_{k=1}^n|a_k|$ is called the weight of the function.
At NNLO accuracy one ends up dealing with harmonic sums of weight up to $w=4$. 

In order to perform the Mellin inverse transformation to $z$-space 
along a contour in the complex $N$ plane at the very end, see Sec.~\ref{subsec:num} below,
one needs to know all functions not only for discrete integers but for any 
complex value of $N$.
This is achieved by proper analytical continuation of the harmonic sums in Eq.~(\ref{eq:harmonicsums}).
As it is well known \cite{ref:cfp}, there is no analytical continuation for all integer values
of $N$ due to the presence of terms $\propto(-1)^N$, and a choice $(-1)^N \to \pm 1$ has to be made based on physical considerations.
For instance, the analytical continuation of all the
coefficient functions $\mathbb{C}_{k,l}^{\mathrm{S,NS}}$ appearing in Eq.~(\ref{eq:nnlostructure}) has to correctly reproduce only
even integer $N$ moments.

To compare our results for the Mellin moments of the NNLO coefficients obtained with the 
help of the {\sc Mathematica} packages \cite{ref:mtpackage, ref:hplpackage}
with those given for even values of $N$ in \cite{ref:nnlo-sia+mellin1},
special care needs to be taken for factors $\propto S_{-2}(N-2)/(N-2)$ since the zero in the denominator 
for $N=2$ suggests the presence of a pole.
However, this is a spurious pole as can be seen by making use of its the integral representation~\cite{ref:harmoinctwoloop}
\begin{eqnarray}
\label{eq:s2}
S_{-2}(N)&=&-\int_0^1 dz \log(z)\frac{(-z)^N-1}{1+z} \, .
\end{eqnarray}
The existence of this spurious pole for $N=2$ at NNLO is the reason for the notation 
adopted in~\cite{ref:nnlo-sia+mellin1}, where the Mellin moments of the coefficient functions 
are written proportional to $\theta(N-3)$ and $\delta(N-2)$,
representing the finite $N\rightarrow2$ limit.
Note that the limit in Eq.~(\ref{eq:s2}) has to be taken for even $N$ to obtain the correct sign.
This can be made manifest by rewriting Eq.~(\ref{eq:s2}) in terms of the digamma function  
which is defined as the derivative of the Euler Gamma function 
$\psi(x) \equiv d \log[\Gamma(x)]/dx$.
The harmonic sum in Eq.~(\ref{eq:s2}) then reads~\cite{ref:ancontpackage}
\be
\label{eq:s22}
S_{-2}(N)= (-1)^{N+3}\beta '(N+1)-\frac{1}{2}\zeta(2)\;,
\ee
where
\begin{equation}
\label{eq:beta}
\beta(N)=\frac{1}{2}\bigg[\psi\left(\frac{N+1}{2}\right)-\psi\left(\frac{N}{2}\right)\bigg]\;.
\end{equation}

We fully reproduce both the $\theta(N-3)$ pieces and the $N\to 2$ limits of the NNLO coefficients
$\mathbb{C}_{k,l}^{\mathrm{S,NS}}(N)$ listed in Ref.~\cite{ref:nnlo-sia+mellin1}.
Note that the subtleties concerning the spurious pole for $N=2$ first appear at the NNLO level.
We also completely agree with the results given in Ref.~\cite{ref:nnlo-sia-mellin2} as long as 
we do not use their definitions of the functions $A_3(N)$, $A_5(N)$, $A_{18}(N)$, $A_{21}(N)$, and $A_{22}(N)$
in Eq.~(14) of \cite{ref:nnlo-sia-mellin2} but, instead, define them as the Mellin transforms of the functions
$g_3(x)$, $g_5(x)$, $g_{18}(x)$, $g_{21}(x)$, and $g_{22}(x)$ specified in 
the {\sc Ancont} package \cite{ref:ancontpackage}.

In our numerical code we implement the Mellin $N$ space expressions for the NNLO coefficient functions 
in the way as they are presented in \cite{ref:nnlo-sia-mellin2}.
The proper analytical continuations of all the harmonic sums and special functions 
are taken from \cite{ref:nnlo-sia-mellin2,ref:ancontpackage,ref:harmoincsums,ref:harmoinctwoloop}. 
In addition, we are making use of some of the routines provided in the {\sc Ancont} package \cite{ref:ancontpackage}.

\subsection{Time-like Evolution Equations \label{subsec:evol}}
%
The factorization procedure invoked in Eq.~(\ref{eq:nnlostructure})
introduces an arbitrary scale $\mu_F$ which conceptually separates the high-energy perturbative regime from 
the low-energy, non-perturbative region.
Both the hard coefficient functions and the FFs depend on $\mu_F$ in such a way that at ${\cal O}(a_s^n)$ in 
pQCD any residual dependence of a physical cross section on $\mu_F$ is of order ${\cal O}(a_s^{n+1})$.
Similar to the case of PDFs, this leads to a set of $2N_f+1$ coupled renormalization group equations (RGE) governing 
the scale $\mu_F$ dependence of the gluon and $N_f$ quark and antiquark FFs into a given hadron species $h$.
Schematically, these time-like evolution equations read
\begin{equation}
\label{eq:evolution}
\frac{\partial}{\partial\ln\mu^2}D^h_i(z,\mu^2)= \sum_j P^T_{ji}(z,\mu^2)\otimes D^h_j\left(z,\mu^2\right)\, ,
\end{equation}
$i,j=q,\bar{q},g$, and where, for simplicity, we have set $\mu_R=\mu_F=\mu$ as in Sec.~\ref{subsec:xsec}.
The $j\to i$ splitting functions $P^T_{ji}(z,\mu^2)$ can be calculated perturbatively as a series in
$a_s$,
\begin{equation}
\label{eq:splittingexp}
P^{T}_{ji} = a_s P_{ji}^{T,(0)} + a_s^2 P_{ji}^{T,(1)} +  a_s^3 P_{ji}^{T,(2)} + \ldots \, ,
\end{equation}
suppressing all arguments $z$, $\mu^2$ in (\ref{eq:splittingexp}).
They are known up to NNLO accuracy \cite{ref:nnlo-kernel}, i.e., ${\cal O}(a_s^3)$, 
as is the case for their space-like counterparts $P^{S}_{ij}$  \cite{ref:space-nnlo-kernel} 
needed for the scale evolution of PDFs.
In fact, there is still a small uncertainty left  
concerning the off-diagonal splitting kernel $P_{qg}^{T,(2)}$
which could not be completely determined by the 
crossing relations to the space-like results employed in \cite{ref:nnlo-kernel}.
Presumably, this remaining ambiguity is numerically irrelevant for all phenomenological applications;
see, however, Ref.~\cite{ref:Pijdirectcalculation} for the status of an ongoing direct calculation
of the NNLO time-like kernels. 

To implement the time-like evolution equations (\ref{eq:evolution}) numerically
up to NNLO accuracy, we closely follow the strategies and framework developed for the public, 
space-like PDF evolution code {\sc Pegasus}~\cite{ref:pegasus}.
In general, the structure and solutions of the space-like and time-like evolution equations are
completely analogous apart from replacing PDFs by FFs and the kernels $P^{S}_{ij}$ by $P^{T}_{ji}$. 
Hence, for completeness, we repeat here only the most important aspects, in particular, those
features appearing for the first time at NNLO.
 
Instead of working directly with the system of $2N_f+1$ coupled equations in (\ref{eq:evolution}) 
it is convenient to recast the quark sector into a flavor singlet 
\be \label{eq:singlet}
D^h_{\Sigma} \equiv \sum_{q}^{N_f}(D^h_{q}+D^h_{\bar q}) \, ,
\ee
which evolves along with the gluon FF $D_g^h$, 
\begin{equation}
\label{eq:singletevol}
\frac{d}{d\ln\mu^2} \Bigg (\begin{matrix}D^h_{\Sigma} \\[2mm] D^h_g
\end{matrix} \Bigg ) =  \Bigg (\begin{matrix}P^T_{qq} & 2N_f P^T_{gq}\\[2mm] \frac{1}{2N_f}P^T_{qg} & P^T_{gg} \end{matrix} \Bigg ) \otimes \Bigg (\begin{matrix}D^h_{\Sigma} \\[2mm] D^h_g
\end{matrix} \Bigg ) \, ,
\end{equation}
and $2N_f-1$ non-singlet combinations 
\ba \label{eq:NSPM}
D^{h,\pm}_{\mathrm{NS},l} &\equiv& \sum_{i=1}^k (D^h_{q_i}\pm D^h_{\bar q_i}) - k (D^h_{q_k}\pm D^h_{\bar q_k})\, , \\ 
\label{eq:NSVAL}
D^h_{\mathrm{NS},v} &\equiv& \sum_{q}^{N_f}(D^h_{q}-D^h_{\bar q})\,,
\ea
reflecting the properties of the (anti)quark to (anti)quark splitting functions and which
all evolve independently. In Eq.~(\ref{eq:NSPM}) $l=k^2-1$, $k=1,\ldots,N_f$, and
the subscripts $i,k$ were introduced to distinguish different quark flavors.
After the evolution is performed, the individual $D_q^h$ and $D_{\bar{q}}^h$ can
be recovered from Eqs.~(\ref{eq:singlet}), (\ref{eq:NSPM}), and (\ref{eq:NSVAL}),
and any combination relevant for a cross section calculation can be computed,
such as those used in the factorized expression for SIA given in Eq.~(\ref{eq:nnlostructure}).

More specifically, the three NS combinations in Eq.~(\ref{eq:NSPM}) and (\ref{eq:NSVAL})
evolve with the following NS splitting functions \cite{ref:nnlo-kernel}
\ba
\label{eq:splittingNSPM}
P^{T,\pm}_{\mathrm{NS}}&=&P^{T,v}_{qq}\pm P^{T,v}_{q\bar q}\,,\\ \nn
P^{T,v}_{\mathrm{NS}}&=& 
\label{eq:splittingNSVAL}
P^{T,-}_{\mathrm{NS}}+P^{T,s}_{\mathrm{NS}} \, ,
\ea
respectively, and the singlet $P_{qq}^T$ in (\ref{eq:singletevol}) obeys
\begin{eqnarray}
\label{eq:splittingS}
P^{T}_{qq} 
= P^{T,+}_{\mathrm{NS}}+P^{T,ps} \, .
\end{eqnarray}
Similarly to the space-like case, 
$P^{T,v}_{q\bar q} = P^{T,s}_{\mathrm{NS}} = P^{T,ps} = 0$ and
$P^{T,s}_{\mathrm{NS}}=0$ in LO and NLO, respectively, such that three independently
evolving NS quark combinations appear for the first time at NNLO accuracy \cite{ref:nnlo-kernel}. 
We note that $P^{T,s}_{\mathrm{NS}}\neq 0$ can lead to a 
perturbatively generated, albeit small strange-quark asymmetry for FFs, i.e.,
$D_s^{h}(z,\mu^2)-D_{\bar s}^{h}(z,\mu^2)\neq 0$,
even if the input $D_s^{h}$ and $D_{\bar{s}}^{h}$ are symmetric;
see Ref.~\cite{Catani:2004nc} 
for a detailed discussion of a similar effect in the context of PDFs.
For pion FFs such a charge asymmetry is expected to
be further suppressed since the effect is driven by a non-zero
$D^h_{\mathrm{NS},v} $ in Eq.~(\ref{eq:NSVAL}).
This combination vanishes when exact charge conjugation and
isospin symmetry is imposed on the $u$ and $d$ quark and
antiquark FFs as is the case in many of the available sets
of pion FFs \cite{ref:kretzer,ref:other-ffs,ref:hkns}.

As mentioned already, we choose to solve the set of time-like evolution equations 
in Mellin $N$ space, which not only has the benefit of turning all integro-differential equation into
ordinary differential equations but also makes them amenable to further analytical studies.
Solutions of the evolution equations in $N$ space, as well as their 
numerical implementation, are well known and were treated extensively in, e.g., Ref.~\cite{ref:pegasus}
in the space-like case relevant for PDFs.
Since the procedure for FFs is essentially the same, we will in the following only
sketch some aspects of the solution at NNLO important for our discussions later on.
The needed NNLO kernels $P_{ji}^{T,(2)}(N)$ can be found in \cite{ref:nnlo-kernel}. 
As for the SIA coefficient functions presented in Sec.~\ref{subsec:xsec}, we have 
verified the expressions for $P_{ji}^{T,(2)}(N)$ starting from $z$-space
and find full agreement.

We start our discussions by recalling the Mellin transformed time-like evolution equations. 
Adopting the notations used in the {\sc Pegasus} code \cite{ref:pegasus}, one finds
\ba
\label{eq:mtevolution}
\frac{\partial \boldsymbol D^h(N,a_s)}{\partial a_s}&=&-\frac{1}{a_s}\bigg[\boldsymbol R_0(N)+\sum^\infty_{k=1}a_s^k\boldsymbol R_k(N)\bigg]\boldsymbol D^h(N,a_s)\, , \nn \\
\ea
where the bold characters indicate that we are dealing in general with $2\times 2$ matrix-valued equations, 
cf. Eq.~(\ref{eq:singletevol}). For the NS combinations (\ref{eq:NSPM}) and (\ref{eq:NSVAL}), 
Eq.~(\ref{eq:mtevolution}) reduces to a set of single partial differential equations
which are straightforward to solve, and we do not consider them here any further.

The $\boldsymbol R_k$ in~(\ref{eq:mtevolution}) are defined recursively as
\be \label{eq:Rmatrix}
\boldsymbol R_0\equiv \frac{1}{\beta_0} \boldsymbol P^{T,(0)}\;,\;\; \boldsymbol R_k \equiv \frac{1}{\beta_0}\boldsymbol P^{T,(k)}-\sum^k_{i=1}b_i\boldsymbol R_{k-i}\;,
\ee
where $\boldsymbol P^{T,(k)}(N)$ is the $k$-th term in the perturbative expansion of the
$2\times 2$ matrix of singlet splitting functions, cf.\ Eq.~(\ref{eq:singletevol}).
In addition, $b_i \equiv \beta_i/\beta_0$ with $\beta_k$ denoting the expansion coefficients of the QCD $\beta$-function;
see Ref.~\cite{ref:betafct} for explicit expressions up to NNLO, i.e, $\beta_2$.
Also note that Eq.~(\ref{eq:mtevolution}) is now written in terms of $\partial a_s$ rather than 
$\partial \log \mu^2$ used in Eq.~(\ref{eq:evolution}).
This convenient change of variables is possible as long as factorization and renormalization scales
are related by a constant, i.e., $\mu_R=\kappa \mu_F$, in numerical studies;
see Ref.~\cite{ref:pegasus} for a detailed discussion. 
For simplicity, we have so far only considered the case $\mu=\mu_R=\mu_F$. Expressions for $\kappa\neq 1$
can be easily recovered both for the coefficient functions (\ref{eq:coeffexp}) 
and the splitting functions (\ref{eq:splittingexp}) by re-expanding $a_s$ in powers of 
$\log (\mu_F^2/\mu_R^2)$. The general expressions are implemented in our numerical code.

Due to the matrix-valued nature of Eq.~(\ref{eq:mtevolution}), no unique closed
solution exists beyond LO. Instead, it can be written as an expansion around the LO solution, 
$(a_s/a_0)^{-\boldsymbol R_0(N)} \boldsymbol D^h(N,a_0)$, where $a_0$ is the value of $a_s$ at the
initial scale $\mu_0$ where the non-perturbative input $\boldsymbol D^h(N,a_0)$ is specified from a fit to data.
This expansion reads
\begin{eqnarray}
\label{eq:mtgeneralsolution}  
\boldsymbol D^h(N,a_s)&=&\bigg[1+\sum^\infty_{k=1}a_s^k\,\boldsymbol U_k(N)\bigg]
\bigg(\frac{a_s}{a_0}\bigg)^{-\boldsymbol R_0(N)} \nonumber\\[2mm]
&\times & \bigg[1+\sum^\infty_{k=1}a_s^k\,\boldsymbol U_k(N)\bigg]^{-1}\boldsymbol D^h(N,a_0)\;.\hspace*{5mm}
\end{eqnarray}
The evolution matrices $\boldsymbol U_k$ are recursively defined by the commutation relations
\begin{eqnarray}
\label{eq:umatrix}
[\boldsymbol U_k,\boldsymbol R_0]=\boldsymbol R_k+\sum_{i=1}^{k-1}\boldsymbol R_{k-1}\boldsymbol U_i+k\boldsymbol U_k\;.
\end{eqnarray}

Based on (\ref{eq:mtgeneralsolution}), it is now possible to define several solutions at order N$^\text m$LO
which are all equivalent up to the accuracy considered, i.e., up to subleading higher-order terms.
Any numerical differences between two different choices should be treated as a source of theoretical uncertainty in
the determination of FFs or PDFs, and it is expected that the inclusion of NNLO corrections reduces this
type of ambiguity as compared to NLO.
We highlight two possible solutions which we pursue further in 
our phenomenological studies in Sec.~\ref{sec:fit}.
Suppose the perturbatively calculable quantities $\boldsymbol P^{T,(k)}$ and $\beta_k$ are 
available up to a certain order $k=m$.
One possibility is to expand Eq.~(\ref{eq:mtgeneralsolution}) in $a_s$ and strictly keep only terms up to $a_s^m$. 
This defines what is usually called the \emph{truncated solution} in Mellin moment space, and, unless stated otherwise,
will be used in all our phenomenological applications.

However, given the iterative definition of the $\boldsymbol R_k$ in Eq.~(\ref{eq:Rmatrix}), 
one may alternatively calculate the $\boldsymbol R_k$ and, 
hence the $\boldsymbol U_k$ in Eq.~(\ref{eq:umatrix}), 
for any $k>m$ from the known results for $\boldsymbol P^{T,(k)}$ and $\beta_k$ up to $k=m$.
Any higher order $\boldsymbol P^{T,(k)}$ and $\beta_k$ with $k>m$ are simply set to zero.
Taking into account all the thus constructed $\boldsymbol U_k$
in Eq.~(\ref{eq:mtgeneralsolution}) defines the so-called \emph{iterated solution}. 
This solution is important as it
mimics the results that would be obtained by solving Eq.~(\ref{eq:evolution}) directly in $z$-space by some numerical
iterative method. 
Both choices are equally valid as they only differ by terms that are of order $\mathcal O (a_s^{m+1})$
and are implemented in our numerical code; see Ref.~\cite{ref:pegasus} for a more detailed discussion 
in the context of space-like evolution equations.
We shall illustrate the numerical differences between the truncated and iterated solution in Sec.~\ref{sec:fit}.

\subsection{Numerical Implementation \label{subsec:num}}
%
We base the development of our new NNLO evolution code for FFs on the well-tested 
{\sc Pegasus} package~\cite{ref:pegasus} which provides different numerical  
solutions to the space-like evolution of PDFs up to NNLO accuracy in Mellin $N$ space
and the necessary routines for the subsequent Mellin inverse transformation back to
momentum space. It also solves the RGE for the strong coupling $a_s(\mu_R^2)$ in the required order in pQCD. 
In addition to extending {\sc Pegasus} to handle also time-like evolution, we also
add packages to compute the SIA cross section in $N$-space and to determine the parameters of the FFs
at some input scale $\mu_0$ from a fit to existing SIA data at LO, NLO, and NNLO accuracy.

In Sec.~\ref{subsec:evol} we have omitted how we deal with heavy quark flavors, i.e.,
charm and bottom, in the time-like scale evolution apart from defining the relevant
$2N_f-1$ NS combinations of FFs in Eqs.~(\ref{eq:NSPM}) and (\ref{eq:NSVAL}).
In {\sc Pegasus}~\cite{ref:pegasus}
both a fixed flavor-number scheme (FFNS) and a variable flavor-number scheme (VFNS) evolution
are implemented. For the latter, matching coefficients between the space-like evolution for
$N_f$ and $N_f+1$ are provided for both PDFs \cite{ref:pdfmatch} and the RGE for $a_s$ \cite{ref:asmatch} up to NNLO accuracy. 
Similar time-like matching coefficients for FFs are only known up to NLO and can be found
in Ref.~\cite{ref:hq-ffs}. They are implemented in our evolution code.
In practice, however, all fits of FFs performed so far \cite{ref:dss,ref:dss2,ref:dssnew,ref:other-ffs,ref:hkns,ref:kretzer},
have used a different approach for the charm and bottom-to-light hadron FFs.
Once the scale $\mu$ in the evolution crosses the heavy quark pole mass $Q=m_{c,b}$, 
a new non-perturbative input distribution is introduced at that scale $D_{c,b}^h(z,m_{c,b}^2)$ and $N_f \to N_f+1$.
The parameters describing these input distributions $D_{c,b}^h(z,m_{c,b}^2)$ are also determined 
by a fit to, usually flavor-tagged, data taken at scales $\mu \gg m_{c,b}$.
We will also adopt this non-perturbative input scheme (NPIS) in all our phenomenological studies below.
We note that as one of the many cross-checks for our new time-like evolution code, we have implemented the 
input parameters and $a_s(\mu_0)$ value of the NLO NPIS fit to SIA data  
performed in Ref.~\cite{ref:kretzer}.
We obtain an excellent numerical agreement with the FFs of~\cite{ref:kretzer} for all $z$ and $\mu$ values.

\begin{figure}[thb!]
\begin{center}
\includegraphics[width=0.4\textwidth]{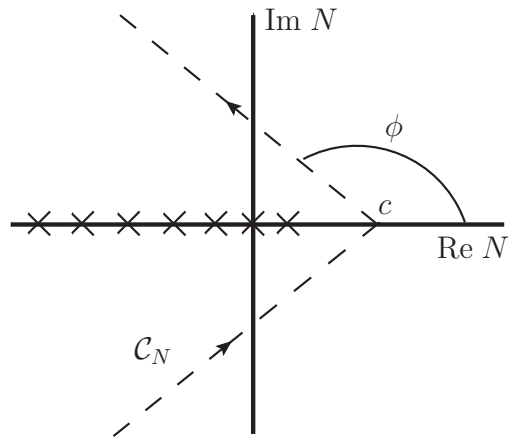}
\end{center}
\caption{The dashed line represents the contour
$\mathcal{C}_N$ in complex $N$-space to perform the inverse Mellin transformation (\ref{eq:inverse1}).
The poles of the integrand along the real axis are schematically represented by the crosses. 
\label{fig:tiltedcontour}
}
\end{figure}
As the last technical issue, we would like to comment on the numerical implementation of the Mellin inverse transformation. 
To this end, one needs to perform a numerical integration in complex $N$-space along a suitably chosen contour 
${\cal C}_N$ in order to recover expressions in $z$-space which can be compared to data.
In case of the SIA cross section, this transformation schematically reads
\be\label{eq:inverse1}
D(z) \otimes \mathbb{C}(z)=\f{1}{2\pi i}\int_{{\cal C}_N} dN\, z^{-N}\, D(N)\,\mathbb{C}\left(N\right) \, ,
\ee
where we have omitted any scale $\mu$ and flavor dependence in Eq.~(\ref{eq:inverse1}). 
In practice, one chooses a tilted contour ${\cal C}_N$ which can be parametrized in terms of a 
real variable $x$ as $N=c+x\, e^{i\phi}$, see Fig.~\ref{fig:tiltedcontour}
for an illustration of the path and Ref.~\cite{ref:pegasus} for more details.
To ensure that the value of the integral is independent of ${\cal C}_N$,
$c$ has to be to the right of the rightmost pole of the integrand, which, in our case, are all 
located along the real axis. An exponential dampening of the integrand in (\ref{eq:inverse1}) is 
achieved for $\pi > \phi \geq \pi/2$, resulting in a smaller upper integration limit $x_{\max}$ 
sufficient for a numerically stable result.

\begin{figure}[thb!]
\vspace*{-1cm}
\begin{center}
\includegraphics[width=0.48\textwidth]{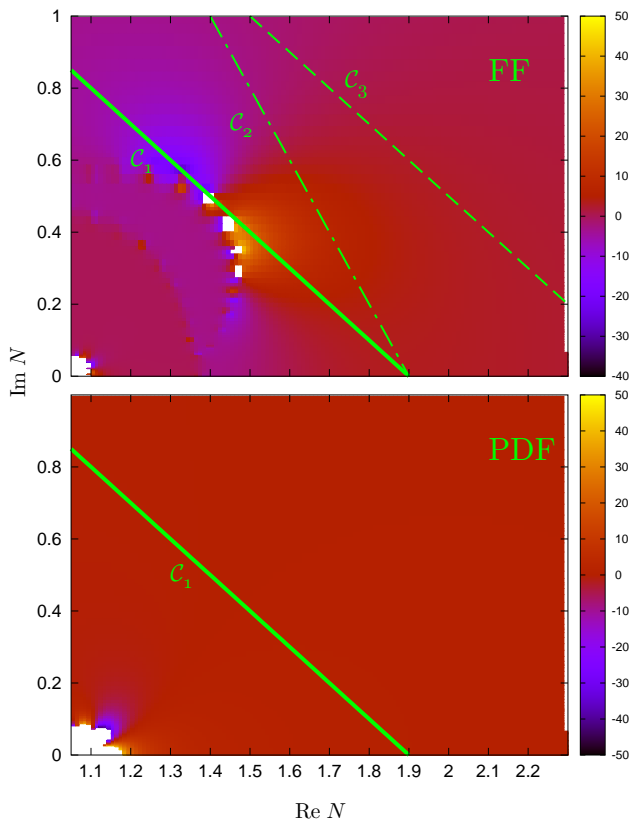}
\end{center}
\vspace*{-1cm}
\caption{The value of the real part of $\mathcal{K}_{12}$ in Eq.~(\ref{eq:mtgeneralsolutionschema}) in a region of the
complex $N$ plane for both the evolution of FFs (upper panel) and PDFs (lower panel).
The lines correspond to three different integration contours ${\cal C}_N$ in (\ref{eq:inverse1}). 
$\mathcal{C}_1$ is the default choice in the {\sc Pegasus} package \cite{ref:pegasus}; see text.
\label{fig:contour} }
\end{figure}
However, extra care needs to be taken in choosing actual values for 
both $c$ and $\phi$ beyond the requirements just outlined.
As it turns out, the standard choice, $c=1.9$ and $\phi=3/4$, made for the PDF evolution in {\sc Pegasus} 
cannot be used in the time-like case.
This is due to the fact that the time-like kernels $P^{T}(z)$ are more singular than their space-like counterparts
$P^{S}(x)$ in the limit $z,x \to 0$.
At NLO accuracy, one finds, for instance, that $P_{gg}^{T,\,(1)}(z)\propto \log^2(z)/z$~\cite{ref:nnlo-kernel}
whereas $P^{S,(1)}_{gg}(x)\propto 1/z$~\cite{ref:space-nnlo-kernel}. In Mellin space this behavior translates into
$\propto 1/(N-1)^3$ and $\propto 1/(N-1)$, respectively, i.e., a leading singularity at $N=1$. 
To order $\mathrm{N^{m}LO}$ this generalizes to $P^{T,\,(m)}_{gg}(N)\propto 1/(N-1)^{(2m+1)}$ \cite{ref:smallz}
whereas in the space-like case only one additional power of $1/(N-1)$ appears in each order \cite{ref:smallx}.
As a result, the function that is integrated in Eq.~(\ref{eq:inverse1}) has 
potentially much stronger oscillations in the vicinity of the pole $N=1$ 
than for the corresponding Mellin inverse transformations for space-like PDFs and
observables, and achieving numerical convergence becomes considerably more delicate.

To illustrate this issue further, we schematically write the general solution
in Eq.~(\ref{eq:mtgeneralsolution}) as
\be
\label{eq:mtgeneralsolutionschema}
\boldsymbol D^h(N,a_s)= \Bigg (\begin{matrix}\mathcal{K}_{11}^{T}(a_s,a_0,N) & \mathcal{K}^{T}_{12}(a_s,a_0,N)\\[2mm] 
\mathcal{K}^{T}_{21}(a_s,a_0,N) & \mathcal{K}^{T}_{22}(a_s,a_0,N) \end{matrix} \Bigg )\boldsymbol D^h(N,a_0)\;,
\ee
where the $\mathcal{K}_{ij}^T$ denote the entries of the $2\times 2$ time-like 
evolution matrix on the right-hand-side of (\ref{eq:mtgeneralsolution}).
A similar equation can be written down for the evolution of PDFs.
 
In Fig.~\ref{fig:contour} we show a comparison of the real part of the 
NLO singlet evolution kernel $\mathrm{Re}\{\mathcal{K}^{T,S}_{12}\}$ 
for the iterated solution for both the evolution of FFs (upper panel) 
and PDFs (lower panel) in the relevant section of the complex $N$ plane.
As an illustrative example, we have chosen $\mu_0^2=1\;\text{GeV}^2$
and $\mu^2=110\;\text{GeV}^2$, the scale relevant for {\sc{Belle}} and {\sc{BaBar}},
in Eq.~(\ref{eq:mtgeneralsolutionschema}).
The line labeled as ${\cal C}_{1}$ represents the standard contour
${\cal C}_N$ implemented in {\sc Pegasus} \cite{ref:pegasus}, and ${\cal C}_{2,3}$ are two
alternative choices.

As can be seen from the upper panel of Fig.~\ref{fig:contour}, 
the contour $\mathcal{C}_1$ with $c=1.9$ and $\phi=3/4$
goes through a region of strong numerical oscillations of $\mathrm{Re}\{\mathcal{K}^{T}_{12}\}$
and, as a consequence, yields numerically unstable results for the integral in Eq.~(\ref{eq:inverse1}). 
Hence, in our code we need to choose either a different angle, e.g., $\phi=2/3$ as in $\mathcal{C}_2$, or 
a different value of $c$, such as $c=2.5$ adopted in $\mathcal{C}_3$. Both choices lead to 
numerically stable and identical results for the Mellin inverse transformation in Eq.~(\ref{eq:inverse1})
for all practical purposes.
Figure~\ref{fig:contour} also shows that no such issue appears for the evolution of PDFs because of the 
weaker $N=1$ singularity than in the time-like case.

Finally, we compare the results of our time-like evolution code with those obtained with the 
publicly available {\sc Mela}~\cite{ref:apfelmela} package, where also tables of benchmark numbers
are given corresponding to input FFs taken from the fit in Ref.~\cite{ref:hkns}; cf.~Eq.~(3.3) 
in \cite{ref:apfelmela}. Using the same input FFs,  
we were not able to directly reproduce their benchmark results 
as generated ``out of the box''from the downloadable script.
The RGE for $a_s(\mu_R)$ is always solved exactly in our code
by means of a fourth order Runge-Kutta integration~\cite{ref:rungekutta}
(as taken from the {\sc Pegasus} package \cite{ref:pegasus}), whereas in 
{\sc Mela} the standard, expanded solution is utilized for the truncated solution
of Eq.~(\ref{eq:mtgeneralsolution}).
After this small difference is accounted for, we achieve perfect numerical agreement with differences of less than $0.01\%$
for both the truncated and iterated solution using the FFNS with $N_f=3$ or the VFNS.

\section{Phenomenological Applications \label{sec:fit}}
%
As a first application of our time-like evolution package presented in Sec.~\ref{sec:nnlo},
we will perform a fit
to the available SIA data with identified pions up to NNLO accuracy in Sec.~\ref{subsec:fit}.
The obtained sets of LO, NLO, and NNLO pion FFs will be used in Sec.~\ref{subsec:appl} to 
demonstrate the relevance of the NNLO corrections to the SIA cross section and to estimate the
residual theoretical uncertainties due to variations of the factorization scale in each order
or to the choice of a truncated or iterated variant of the solution 
to the evolution equations given in (\ref{eq:mtgeneralsolution}).

\subsection{Fit of Pion FFs up to NNLO Accuracy \label{subsec:fit}}
%
Since full NNLO corrections are only available for a rather limited set of hard scattering processes,
we have to restrict our first analysis of FFs at NNLO accuracy to data obtained in SIA for the time being.
In addition, we focus solely on pion production where data are most abundant and precise. In any case,
the main interest of this work are the general features of NNLO corrections rather than to provide a
new set of FFs. 

To facilitate the fitting procedure, we closely follow the framework outlined and used in the series of DSS
global QCD analyses of parton-to-pion FFs at NLO accuracy \cite{ref:dssnew,ref:dss,ref:dss2}.
Specifically, we adopt the same flexible functional form
\be\label{eq:Dparam}
D_i^{\pi^{+}}(z,\mu_0^2)=\f{N_i\, z^{\alpha_i}(1-z)^{\beta_i}[1+\gamma_i(1-z)^{\delta_i}]}
{B[2+\alpha_i,\beta_i+1]+\gamma_i B[2+\alpha_i,\beta_i+\delta_i+1]} 
\ee
to parametrize the non-perturbative input FFs for charged pions
at a scale $\mu_0$ in the $\overline{\mathrm{MS}}$ scheme.
Here, $B[a,b]$ is the Euler Beta function used to normalize the parameter $N_i$ in (\ref{eq:Dparam})
for each flavor $i$ to its contribution to the energy-momentum sum rule. 
In addition to the gluon $i=g$, we only consider FFs for the sum of a quark and an antiquark of a given flavor $i$, i.e.,
$i = u+\bar u$, $d+\bar{d}$, $s+\bar s$, $c+\bar c$, and $b+\bar b$,
since SIA is only sensitive to $q+\bar q$ flavor combinations as can be already inferred from Eq.~(\ref{eq:nnlostructure}).
Also, since all hadrons in SIA originate from the initially produced $q\bar{q}$ pair, the rates for $\pi^+$ and
$\pi^-$ are the same, and data for charged pions are usually presented for the sum 
$d\sigma^{\pi} \equiv d\sigma^{\pi^{+}}+d\sigma^{\pi^{-}}$.

We assume charge conjugation and isospin symmetry and impose $D_{u+\bar{u}}^{\pi^{\pm}} = D_{d+\bar{d}}^{\pi^{\pm}}$
as is also suggested by the flavor composition of $\pi^{\pm}$. 
We note that a recent global QCD analysis of pion FFs at NLO accuracy based on SIA, SIDIS, and $pp$ data \cite{ref:dssnew}
finds a breaking of this symmetry of less than $0.5\%$.
Beyond that, we are forced to fix certain parameters in our ansatz (\ref{eq:Dparam}) as they cannot be constrained by
data. More specifically, we set $\alpha_{s+\bar{s}} = \alpha_{u+\bar{u}}$,
$\beta_{s+\bar{s}} = \beta_{u+\bar{u}} + \delta_{u+\bar{u}}$, and $\beta_g=8$.
In addition, $\delta_{g,s+\bar{s},c+\bar{c}} = 0$ and $\gamma_{g,s+\bar{s},c+\bar{c}} = 0$.
For light quark flavors and the gluon, we choose an initial scale of $\mu_0=1$~GeV.
As in all previous fits \cite{ref:dss,ref:dss2,ref:dssnew,ref:other-ffs,ref:hkns,ref:kretzer},
the charm and bottom-to-pion FFs are treated as a non-perturbative input
and are turned on discontinuously at $\mu_0^c=m_c=1.4$~GeV and $\mu_0^b=m_b=4.75$~GeV, respectively.
Their parameters are essentially determined by charm and bottom flavor-tagged SIA data. In case of
$D_{b+\bar{b}}^{\pi^{+}}$, a good fit is only achieved with the full functional form (\ref{eq:Dparam})
using all five parameters, whereas for charm only three free parameters are needed.
Since the heavy quark masses are neglected throughout in the NPIS, $D_{c+\bar{c}}^{\pi^{+}}$
and $D_{b+\bar{b}}^{\pi^{+}}$ should be only used in cross sections such as Eq.~(\ref{eq:nnlostructure})
at scales well beyond their partonic thresholds $\mu=2m_c$ and $\mu=2m_b$, respectively.

%
\begin{table}[th!]
\caption{\label{tab:fitpara} Parameters describing our optimum LO, NLO, and NNLO 
$D_i^{\pi^{+}}(z,\mu_0)$ in Eq.~(\ref{eq:Dparam}) at the input scale $\mu_0=1$~GeV.
Results for the charm and bottom FFs refer to the scale 
$\mu_0^c=m_c=1.4$~GeV and $\mu_0^b=m_b=4.75$~GeV, respectively.
The parameters given in italics are fixed by
$\alpha_{s+\bar{s}} = \alpha_{u+\bar{u}}$, 
$\beta_{s+\bar{s}} = \beta_{u+\bar{u}} + \delta_{u+\bar{u}}$,
and $\beta_g=8$ but are listed for completeness.}
\begin{ruledtabular}
\begin{tabular}{cccc}
parameter             & LO     & NLO    & NNLO \\
\hline
$N_{u+\bar{u}}$       & 0.735  & 0.572  & 0.579 \\
$\alpha_{u+\bar{u}}$  & -0.371 & -0.705 & -0.913 \\
$\beta_{u+\bar{u}}$   & 0.953  & 0.816  & 0.865 \\
$\gamma_{u+\bar{u}}$  & 8.123  & 5.553  & 4.062 \\
$\delta_{u+\bar{u}}$  & 3.854  & 1.968  & 1.775 \\
\hline
$N_{s+\bar{s}}$       & 0.243  & 0.135  & 0.271 \\
$\alpha_{s+\bar{s}}$  & \em -0.371 & \em -0.705  & \em -0.913  \\
$\beta_{s+\bar{s}}$   & \em 4.807 & \em 2.784 & \em 2.640 \\
\hline
$N_{g}$               & 0.273 & 0.211  & 0.174 \\
$\alpha_{g}$          & 2.414 & 2.210  & 1.595 \\
$\beta_{g}$           & \em 8.000 & \em 8.000 & \em 8.000 \\
\hline
$N_{c+\bar{c}}$       & 0.405  & 0.302  & 0.338 \\
$\alpha_{c+\bar{c}}$  & -0.164 & -0.026 & -0.233 \\
$\beta_{c+\bar{c}}$   & 5.114  & 6.862  & 6.564 \\
\hline
$N_{b+\bar{b}}$       & 0.462  & 0.405  & 0.445 \\
$\alpha_{b+\bar{b}}$  & -0.090 & -0.411 & -0.695 \\
$\beta_{b+\bar{b}}$   & 4.301  & 4.039  & 3.681 \\
$\gamma_{b+\bar{b}}$  & 24.85  & 15.80  & 11.22 \\
$\delta_{b+\bar{b}}$  & 12.25  & 11.27  & 9.908 \\
\end{tabular}
\end{ruledtabular}
\end{table}
The remaining 16 free parameters are determined by a standard $\chi^2$ minimization procedure
as described, for instance, in Ref.~\cite{ref:dssnew}. They are listed in Tab.~\ref{tab:fitpara}
for our LO, NLO, and NNLO sets of pion FFs.
For each set of experimental data we determine the optimum normalization shift analytically and
assign an additional contribution to $\chi^2$ according to the quoted experimental uncertainties;
see, e.g., Eq.~(5) in Ref.~\cite{ref:dssnew} for details.

\begin{table}[th!]
\caption{\label{tab:exppiontab} The individual $\chi^2$-values and number of points
for each inclusive and flavor-tagged data set included in our fits at LO, NLO, and NNLO accuracy. At the bottom, 
we list the total $\chi^2$-penalty from the normalization
shifts and the total $\chi^2$for each fit.}
\begin{ruledtabular}
\begin{tabular}{lccccc}
experiment& data & \# data & $\chi^2$  &  &\\
          & type & in fit  &  LO & NLO & NNLO       \\\hline
{\sc Sld} \cite{ref:slddata}  
          & incl.\            & 23 &15.0 &14.8 &15.5 \\
          & $uds$ tag         & 14 &9.7 &18.7 &18.8 \\
          & $c$ tag           & 14 &10.4 &21.0 &20.4 \\
          & $b$ tag           & 14 &5.9 &7.1  &8.4  \\
{\sc Aleph} \cite{ref:alephdata}    
          & incl.\            & 17 &19.2 &12.8 &12.6 \\
{\sc Delphi} \cite{ref:delphidata}  
          & incl.\            & 15 &7.4 &9.0 &9.9 \\
          & $uds$ tag         & 15 &8.3 &3.8 &4.3 \\
          & $b$ tag           & 15 &8.5 &4.5 &4.0 \\
{\sc Opal} \cite{ref:opaldata}  
          & incl.\            & 13 &8.9 &4.9 &4.8 \\
{\sc Tpc} \cite{ref:tpcdata}  
          & incl.\            & 13 &5.3 &6.0 &6.9 \\
          & $uds$ tag         &  6 &1.9 &2.1 &1.7 \\
          & $c$ tag           &  6 &4.0 &4.5 &4.1 \\
          & $b$ tag           &  6 &8.6 &8.8 &8.6 \\
{\sc BaBar} \cite{ref:babardata}     
          & incl.\            & 41 &108.7 &54.3 &37.1 \\ 
{\sc Belle} \cite{ref:belledata}     
          & incl.\            & 76 &11.8 &10.9 &11.0\\    \hline 
{\sc norm.\ shifts}
          &                   &    &7.4  & 6.8    & 7.1  \\\hline\hline
{\bf TOTAL:} 
          &                   & 288 &241.0 &190.0 &175.2\\
\end{tabular}
\end{ruledtabular}
\end{table}
Our fits are performed to the following sets of inclusive and flavor-tagged SIA data with identified pions:
{\sc Sld}~\cite{ref:slddata}, {\sc Aleph}~\cite{ref:alephdata}, {\sc Delphi}~\cite{ref:delphidata}, 
and {\sc Opal}~\cite{ref:opaldata}, all taken at a c.m.s.\ energy of $\sqrt{S}=91.2$~GeV, 
{\sc Tpc}~\cite{ref:tpcdata} at $\sqrt{S}=29$~GeV, and 
{\sc BaBar}~\cite{ref:babardata} and {\sc Belle}~\cite{ref:belledata} both at $\sqrt{S}=10.5$~GeV.
The {\sc Sld}, {\sc Delphi} and {\sc Tpc} experiments not only provide 
inclusive SIA measurements but also $uds$, charm and bottom-tagged data. 
All these sets were also used in the recent global analysis presented in Ref.~\cite{ref:dssnew}.

As is customary \cite{ref:dss,ref:dss2,ref:dssnew,ref:other-ffs,ref:hkns,ref:kretzer},
we do not include any data below a certain $z_{\min}$ in the fit 
where finite, but neglected hadron mass effects $\propto M_{\pi}/(z^2 S)$ might become
relevant~\cite{ref:hmc+res}, 
and potentially large logarithmic contributions $\propto \log{z}$, briefly mentioned in Sec.~\ref{subsec:num}, 
need to be resumed to all orders \cite{ref:smallz,ref:smallz-old,ref:smallz-resum}. For all our fits, we choose $z_{\min}=0.075$.
In addition, we employ an upper cut of $z<z_{\max}=0.95$. In this region threshold logarithms $\propto \log(1-z)$ in the coefficient
functions are expected to become increasingly relevant, and, again, all-order resummations are needed
\cite{ref:resum,ref:hmc+res}.
Resummations are rather straightforward to implement in Mellin $N$-space, and, hence, we plan to extend our 
code further by including them based on the knowledge that can be gathered from 
all the available fixed order results at NNLO accuracy for both $z\to 0$ and $z\to 1$
in a dedicated future work.

We note that we are not fitting the initial value $a_s$ at some reference scale in order 
to solve the RGE governing the running of the strong coupling but rather adopt the following boundary conditions
$\as(M_Z)=0.135$ at LO, $\as(M_Z)=0.120$ at NLO, and $\as(M_Z)=0.118$ at NNLO accuracy from the recent
MMHT global analysis of PDFs; see the first reference in \cite{ref:pdfnnlo}. 

\begin{figure*}[thb!]
\vspace*{-1cm}
\begin{center}
\includegraphics[width=0.93\textwidth]{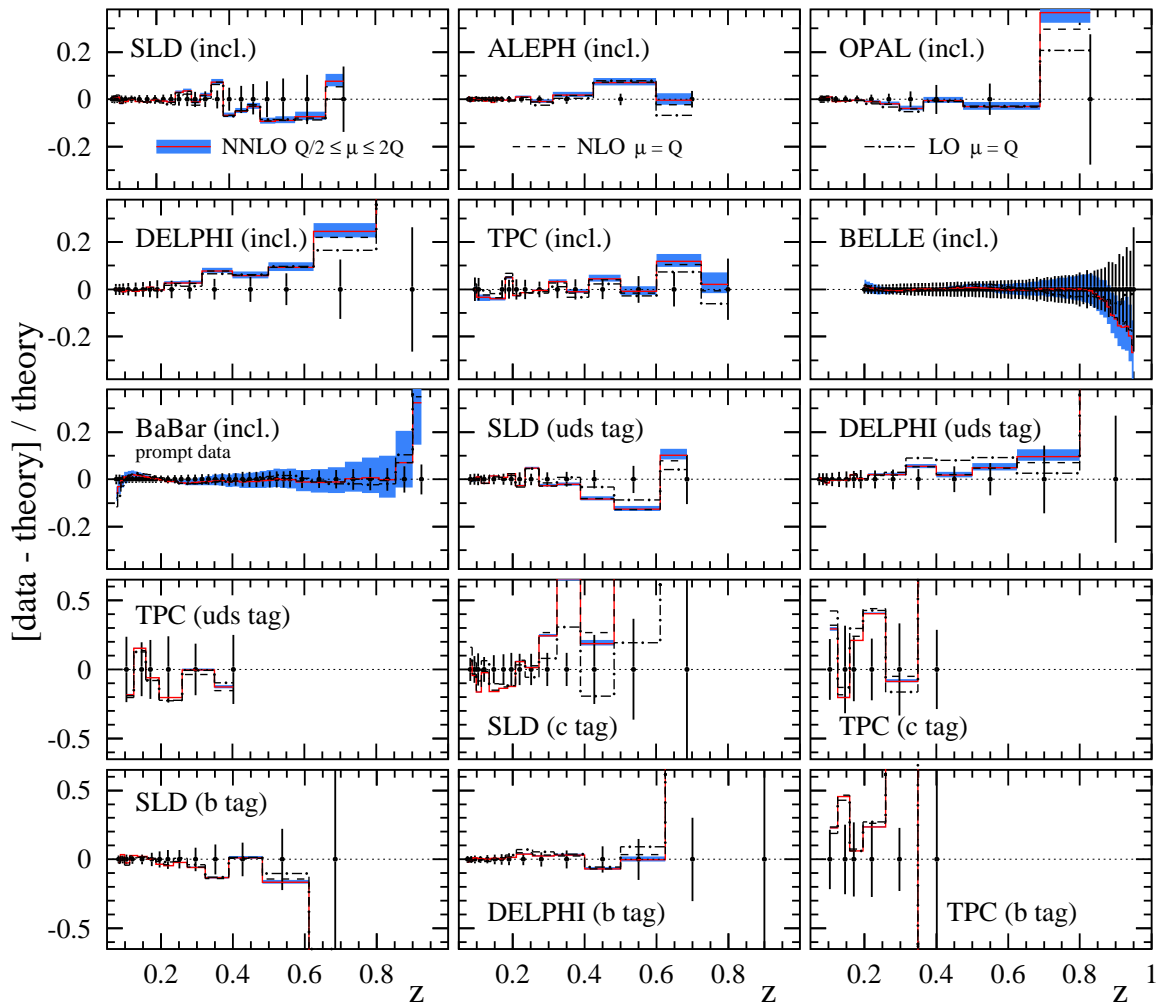}
\end{center}
\vspace*{-1cm}
\caption{Ratios for [data-theory]/theory for our LO (dot-dashed), NLO (dashed), and NNLO (solid lines) 
fits computed with the scale $\mu=Q$ for the data sets listed in Tab.~\ref{tab:exppiontab}.
The shaded bands illustrate the remaining scale ambiguity at NNLO accuracy 
in the range $Q/2\le\mu \le2Q$. The points along the zero axis indicate the relative experimental uncertainty.
\label{fig:thoverdata} }
\end{figure*}

Table~\ref{tab:exppiontab} and Fig.~\ref{fig:thoverdata} illustrate the quality of our fits to SIA data
at LO, NLO, and NNLO accuracy in terms of the individual $\chi^2$-values 
obtained for each experiment and the quantity ``[data-theory]/theory'', respectively.
The total $\chi^2$-penalty originating from the normalization shifts applied to each data set 
can be also found at the bottom of Tab.~\ref{tab:exppiontab}. It turns out to be small, 
about 7 units, and is largely independent of the perturbative order.
Upon applying the cuts on the $z$-range discussed above, 
a total of 288 data points remains for the fitting procedure and to determine the 16
free parameters describing our parton-to-pion FFs $D_{i}^{\pi^{+}}(z,\mu_0)$ in
Eq.~(\ref{eq:Dparam}). All fits yield a very good $\chi^2$ per degree of freedom (d.o.f.)
ranging from 0.89 in LO to 0.64 at NNLO accuracy. We note, however, that the $\chi^2/{\mathrm{d.o.f.}}$
would deteriorate very significantly if the number of free fit parameters would be reduced further
by setting, for instance, $\gamma_{u+\bar{u}}=0$ or $\gamma_{b+\bar{b}}=0$.

As can be seen from Tab.~\ref{tab:exppiontab} and Fig.~\ref{fig:thoverdata}, nearly all SIA data sets can be described 
equally well in LO, NLO, and NNLO accuracy with just a few exceptions, most notably the {\sc BaBar} data \cite{ref:babardata}
taken at the smallest $\sqrt{S}$ which drive the differences found in the total $\chi^2$-values of the three fits. 
Here, the inclusion of higher order corrections progressively leads to better fits.
A closer inspection reveals that the larger $\chi^2$ at LO, and also at NLO, 
stems from the data points corresponding to the lowest $z$ values included in the fit, i.e., $0.075\le z \lesssim 0.12$;
note that the {\sc Belle} Collaboration does not provide any data below $z=0.2$.
This result is readily understood from the fact that calculations at higher orders contain more of
the numerically important small $z$ enhancements $\propto \log z$ mentioned above, i.e., are closer to an all-order result.
From the observation that calculations at NNLO accuracy provide a significantly 
better description of data at small $z$, one can anticipate that including all-order resummations into the 
analysis framework would eventually further extend the range of $z$ amenable to pQCD. We will investigate this
quantitatively in a future publication.
The neglected hadron mass is another source of potentially large corrections at small $z$ and/or $\sqrt{S}$. 
In Ref.~\cite{ref:hmc+res} it was shown, however, that hadron mass terms are relatively small for
pion production in SIA in the kinematic regime relevant for the {\sc BaBar} data.
We also wish to recall that {\sc BaBar} provides their data in two variants called ``conventional'' and ``prompt'',
differing by the treatment of weak decays into pions in their event sample \cite{ref:babardata}.
As in the recent global NLO analysis \cite{ref:dssnew}, our results are based on the latter set. 
We have verified that a decent fit to all SIA data can be also obtained when the ``conventional'' data 
are used instead but at the expense of a less favorable total $\chi^2$, e.g., 236.4 rather than 190.0 units at NLO,
and, more importantly, for undesirable corners of the parameter space describing the $D_{i}^{\pi^{+}}(z,\mu_0)$ in
Eq.~(\ref{eq:Dparam}). For instance, the $u+\bar{u}$ fragmentation tends to saturate the energy-momentum sum rule,
which is summed over all hadrons, already for pions.

Table~\ref{tab:exppiontab} and Fig.~\ref{fig:thoverdata} also reveal that some flavor-tagged data from {\sc Sld} can be described best at LO
but at the expense of larger $\chi^2$-values for inclusive {\sc Aleph} and {\sc Opal} data.
In general, the NLO and NNLO results are very similar for all data sets used in the fits 
except, as just discussed, for a few points from {\sc BaBar} at small $z$. 
This observation also carries over to the obtained FFs at NLO and NNLO accuracy, in particular, those flavor 
combinations which are constrained best by the SIA data alone.

%
\begin{figure*}[th!]
\vspace*{-1.0cm}
\begin{center}
\includegraphics[width=0.6\textwidth]{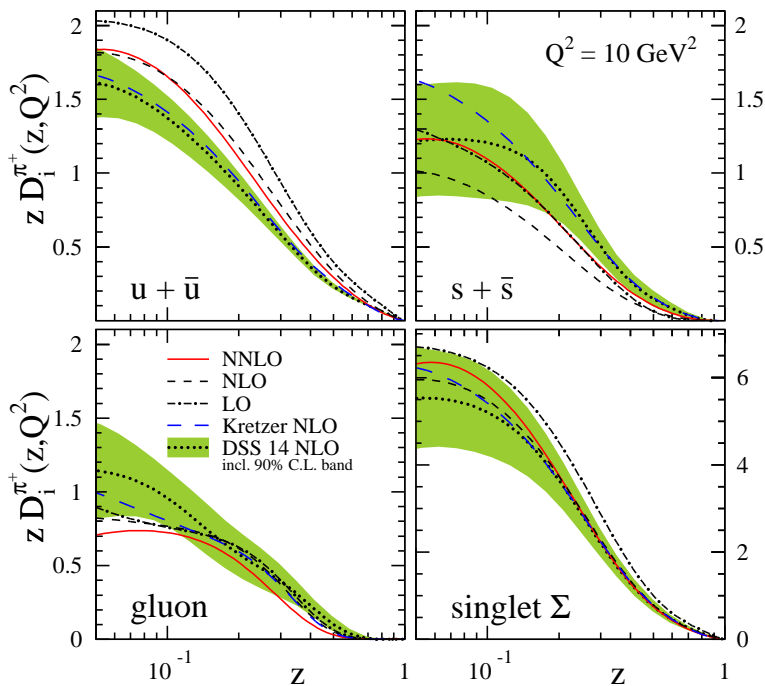}
\end{center}
\vspace*{-1cm}
\caption{Comparison of our LO, NLO, and NNLO FFs $D_{i}^{\pi^{+}}(z,Q^2)$ at $Q^2=10\,\mathrm{GeV}^2$
for $i = u+\bar{u}$, $s+\bar{s}$, $g$, and the flavor singlet combination in (\ref{eq:singlet}) for $N_f=4$.
Also shown are the optimum NLO FFs from Kretzer \cite{ref:kretzer}, obtained also solely from SIA data, and
the latest global analysis of the DSS group \cite{ref:dssnew} based on SIA, SIDIS, and $pp$ data. 
For the latter, we also illustrate their
$90\%$ C.L.\ uncertainty estimates (shaded bands).
\label{fig:DSScomparison} }
\end{figure*}

Figure~\ref{fig:DSScomparison} shows our fitted LO, NLO, and NNLO $D_{i}^{\pi^{+}}(z,Q^2)$ 
at $Q^2=10\,\mathrm{GeV}^2$ for $i = u+\bar{u}$, $s+\bar{s}$, $g$, and the flavor singlet combination 
in (\ref{eq:singlet}) for $N_f=4$. As a comparison with previous NLO results, we consider the most recent global analysis of the DSS group~\cite{ref:dssnew},
based on the same set of SIA data plus SIDIS and $pp$ data, 
and the old fit by Kretzer \cite{ref:kretzer}.
The latter still provides a good description of all pion data, including those from SIDIS and $pp$, despite making use of
only a small subset of the SIA data listed in Tab.~\ref{tab:exppiontab} comprising {\sc Sld} \cite{ref:slddata},
{\sc Aleph} \cite{ref:alephdata}, and {\sc Tpc} \cite{ref:tpcdata}.
To illustrate how the current experimental uncertainties typically propagate to the extraction of parton-to-pion FFs, we also
show in Fig.~\ref{fig:DSScomparison} the $90\%$ confidence level (C.L.) estimates of the latest DSS global QCD fit  (shaded bands).
As was already mentioned, we refrain from providing uncertainty bands for our fits as SIA data alone are not
sufficient for providing a reliable estimate due to the assumptions one has to impose on the parameter space 
describing the $D_{i}^{\pi^{+}}(z,\mu_0)$ in Eq.~(\ref{eq:Dparam}).

From Fig.~\ref{fig:DSScomparison} one can make the following observations:
the quantity which is known to be constrained best by the SIA data alone 
\cite{ref:dss,ref:dss2,ref:dssnew,ref:other-ffs,ref:hkns,ref:kretzer},
the flavor singlet combination $D_{\Sigma}^{\pi^{+}}$ defined in Eq.~(\ref{eq:singlet}), 
is very similar for all the NLO results, DSS, Kretzer, and our fit,
in particular, for $z\gtrsim 0.1$. The fact that also the singlet FF determined at NNLO
accuracy is close to the NLO results gives some indication that NNLO corrections do not seem
to alter results obtained at NLO accuracy too much.
A similar level of agreement for $D_{\Sigma}^{\pi^{+}}$ is found also at other scales, 
for instance, $\mu=M_Z$.

Breaking up the singlet into FFs for individual quark flavors depends on the assumptions made
in the fit, including such details as the choice for $z_{\min}$.
Therefore, it is not too surprising that one finds some differences between the various
fits shown in Fig.~\ref{fig:DSScomparison}
for the favored $D_{u+\bar{u}}^{\pi^{+}}$ and the unfavored $D_{s+\bar{s}}^{\pi^{+}}$,
with the latter FF, of course, being considerably less well constrained by data than the former.
Another FF which is only loosely constrained by a fit to solely SIA data is the gluon $D_{g}^{\pi^{+}}$,
which, despite the different assumptions, agrees rather well among all fits.
Finally, one notices that for a LO fit both the singlet and the favored FFs, 
$D_{\Sigma}^{\pi^{+}}$ and $D_{u+\bar{u}}^{\pi^{+}}$, respectively, are significantly
larger than the corresponding NLO estimates. 
In general, we find that in order to achieve a good fit to SIA data at LO accuracy, 
some of the parameters listed in Tab.~\ref{tab:fitpara}
tend to approach extreme values, for instance, the $u+\bar{u}$ fragmentation  
saturates most of the energy-momentum sum rule already for pions.
In any case, LO estimates are not sufficient for phenomenological applications.

\subsection{Impact of NNLO Corrections on Theoretical Uncertainties \label{subsec:appl}}
%
In this Section we analyze the relevance of the NNLO corrections for a reliable 
phenomenology of the SIA process. To this end, we will examine the importance of 
various sources of theoretical uncertainties in LO, NLO, and NNLO accuracy.
We will present results for the size of the NNLO corrections in terms of the $K$-factor, study
the residual dependence on the choice of scale $\mu$, and
investigate the uncertainties induced by choosing a particular solution, truncated or iterated,
to the time-like evolution equations. 
All these results are largely independent of the details of fitting an actual set of FFs, 
and as such they represent the main numerical results of this paper along with
our newly developed NNLO code described in Sec.~\ref{sec:nnlo}.

\begin{figure}[bh!]
\vspace*{-0.9cm}
\begin{center}
\includegraphics[width=0.49\textwidth]{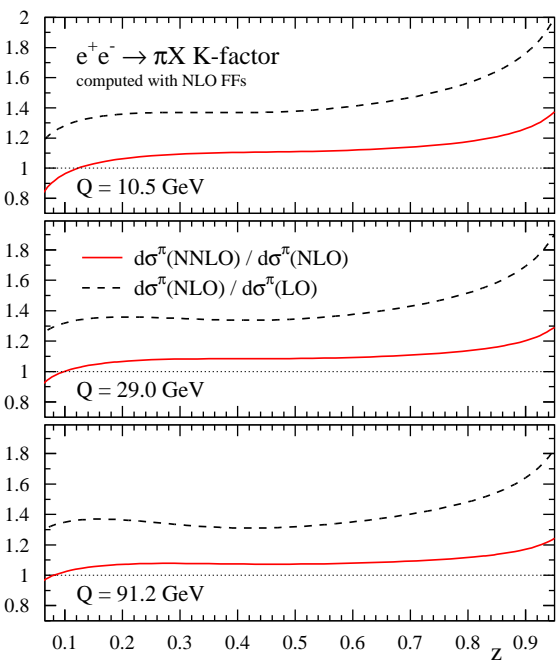}
\end{center}
\vspace*{-0.7cm}
\caption{NNLO/NLO (solid) and NLO/LO (dashed lines) $K$-factors for the SIA process 
for three different c.m.s.\ energies. All computations are performed with our NLO set
of parton-to-pion FFs; see text.
\label{fig:kfactor} }
\end{figure}
%
%
\begin{figure}[tbh!]
\vspace*{-1cm}
\begin{center}
\includegraphics[width=0.5\textwidth]{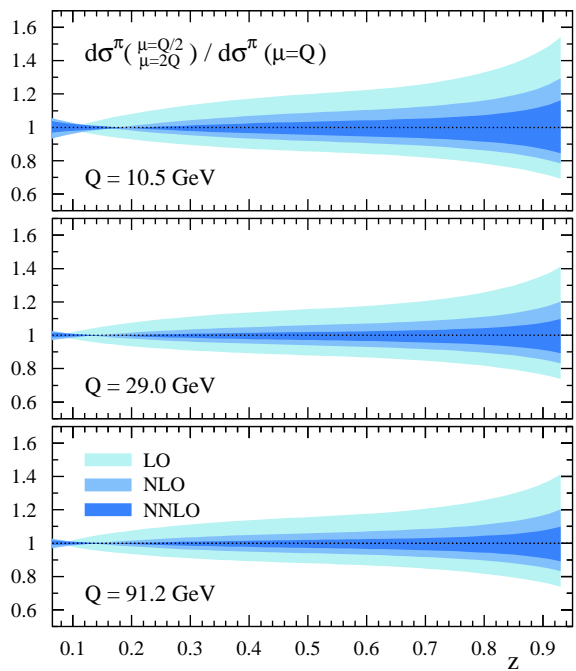}
\end{center}
\vspace*{-1cm}
\caption{Scale dependence of the SIA cross section at LO, NLO, and NNLO accuracy 
in the range
$Q/2 \le \mu=\mu_R=\mu_F \le 2Q$ normalized to the results obtained for $\mu=Q$
for three values of $\sqrt{S}$.
\label{fig:scaledependence} }
\end{figure}
In Fig.~\ref{fig:kfactor}, we show the $K$-factor for the SIA process defined as
$d\sigma^{\pi}\text{(N}^{\text{m}}\text{LO)}/d\sigma^{\pi}\text{(N}^{\text{m-1}}\text{LO)}$
for $m=2$ (solid) and $m=1$ (dashed lines) for the three c.m.s.\ energies 
corresponding to the experiments included in our fit; see Tab.~\ref{tab:exppiontab}.
To determine only the impact of the genuine higher order corrections and not some 
numerical differences in the LO, NLO, and NNLO FFs, like those illustrated in Fig.~\ref{fig:DSScomparison},
all calculations in Fig.~\ref{fig:kfactor} are performed with our NLO input FFs. 
Their evolution, the running of the strong coupling $a_s$, and the coefficient functions are
taken consistently either at LO, NLO, or NNLO accuracy though.

As one expects, the $K$-factor for the NNLO/NLO results is significantly smaller than the one 
for NLO/LO, and for most values of $z$ the additional NNLO corrections are at the level of about
$10\%$ or less. Both at large and small $z$, one finds clear indications for the presence of
large logarithmic corrections to the perturbative series contained in the evolution kernels
$P^T$ and the SIA coefficient functions $\mathbb{C}$. They need to be resummed to all orders
to extend the range of applicability of the presented fixed order results to both $z\to 1$ and
$z\to 0$.
We note that the small $\sqrt{S}$ dependence of the $K$-factors in Fig.~\ref{fig:kfactor} 
is only caused by the different orders in pQCD used in the denominator and in the numerator, 
$d\sigma^{\pi}\text{(N}^{\text{m}}\text{LO)}$ and
$d\sigma^{\pi}\text{(N}^{\text{m-1}}\text{LO)}$, respectively, 
to compute the scale evolution of FFs and the coupling $a_s$.
There is no scale in the coefficient functions as we have set $\mu_R=\mu_F=\mu=Q$
throughout, i.e., all
logarithms of the type $\log(\mu_R^2/\mu_F^2)$ or $\log(Q^2/\mu_F^2)$ vanish.

\begin{figure}[th!]
\vspace*{-1cm}
\begin{center}
\includegraphics[width=0.5\textwidth]{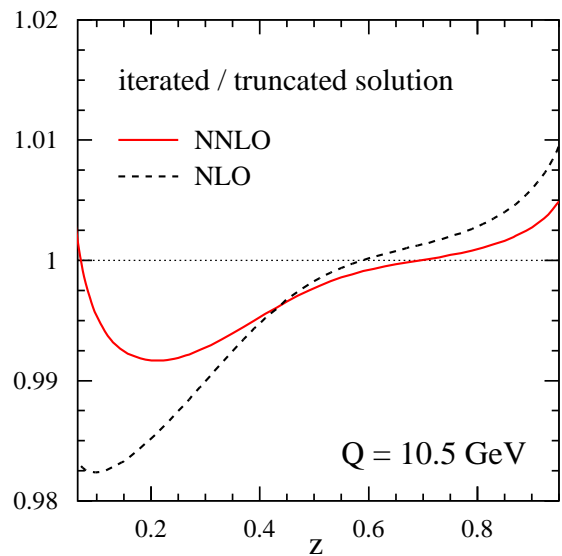}
\end{center}
\vspace*{-1cm}
\caption{Ratio of the iterated and truncated variant of the solution (\ref{eq:mtgeneralsolution})
to the time-like evolution equations at NLO (dashed) and NNLO (solid line) accuracy at the scale of the {\sc BaBar} and {\sc Belle} 
experiments. \label{fig:itovertrunc} }
\end{figure}
The scale dependence of the SIA cross section 
is illustrated in Fig.~\ref{fig:scaledependence}, where we show results at LO, NLO, and NNLO
accuracy (shaded bands) for $\mu_R=\mu_F=\mu=2Q$ and $\mu=Q/2$ normalized in each case to our default
choice $\mu=Q$. 
The residual dependence on the choice of the scale $\mu$ in a theoretical calculation
is presumably the most important source of uncertainty and is expected to shrink progressively
upon including higher and higher order corrections.
This is exactly what we find. For instance, at $\sqrt{S}=10.5\,\mathrm{GeV}$, relevant
for {\sc BaBar} and {\sc Belle}, the typical scale uncertainty at $z\approx 0.5$ amounts to 
about $20\%$ at LO and reduces to $\approx 10\%$ at NLO and $\approx 5\%$ at NNLO.
At larger c.m.s.\ energies, the scale ambiguities are even smaller and 
reach around $1-2\%$ at NNLO accuracy. This is actually needed in a phenomenological analysis 
to roughly match the experimental uncertainties for the most precise sets of inclusive pion data as can be inferred 
from Fig.~\ref{fig:thoverdata}; note that the scale uncertainty bands are hardly visible for some
of the flavor-tagged data as we had to inflate the axis of the ordinate in Fig.~\ref{fig:thoverdata}
to accommodate the rather sizable experimental uncertainties.

As can be seen from Fig.~\ref{fig:scaledependence}, all scale uncertainty bands narrow down somewhere
in the range $0.1\lesssim z \lesssim 0.15$ before they start to increase again towards $z\to 0$.
This can be readily understood from fact that one has approximate ``scaling'' of the SIA cross section,
or, alternatively, the quark FFs, for some value of $z$ in this region, i.e., they become independent of the
scale $\mu$. This is very much similar to DIS and PDFs, where this happens somewhere near
momentum fractions of about 0.2. Of course, QCD corrections always introduce some scale dependence,
and higher order cross sections never probe a FFs or a PDFs locally at one value of momentum fraction 
but rather over a broad range due to the presence of convolutions, like, for instance, in Eq.~(\ref{eq:nnlostructure}).

We close our discussions about the relevance of the NNLO corrections by showing the theoretical ambiguity
associated with the different choices one has in defining the solution to the time-like evolution
equations beyond the LO accuracy. More specifically, Fig.~\ref{fig:itovertrunc} gives the ratio of the 
iterated and truncated variant of the general solution given in Eq.~(\ref{eq:mtgeneralsolution})
computed in NLO (dashed) and NNLO (solid line); see also the corresponding discussions in Sec.~\ref{subsec:evol}.
In the $z$-range relevant for the extraction of FFs from data, this type of theoretical uncertainty is rather small,
and we note that it is usually not considered or even mentioned \cite{ref:dss,ref:dss2,ref:dssnew,ref:other-ffs,ref:hkns,ref:kretzer}.
As for the $K$-factor and the scale dependence shown in Fig.~\ref{fig:kfactor}
and \ref{fig:scaledependence}, respectively, including NNLO corrections reduces the residual uncertainties by about a
factor of two as compared to the results obtained at NLO accuracy. 
For most values of $z$, the differences between the truncated and iterated solutions are less than $0.5\%$ at NNLO, i.e.,
smaller than scale uncertainties and potentially missing higher order corrections as indicated by the $K$-factor
for NNLO/NLO.

\section{Conclusions and Outlook \label{sec:conclusions}}
%
We have presented a first analysis of parton-to-pion fragmentation functions at next-to-next-to-leading order
accuracy in QCD based on single-inclusive pion production in electron-positron annihilation.
To this end, we have extended the existing space-like evolution package {\sc Pegasus} for parton distribution
functions to the time-like region and fragmentation functions. 
The code is numerically very efficient and works throughout in Mellin $N$ moment space, 
where the evolution equations can be solved analytically. 

We have discussed all the relevant technical details to perform the QCD
scale evolution and cross section calculation for single-inclusive hadron production 
in electron-positron annihilation up to next-to-next-to-leading order
accuracy. We have verified all the needed expressions for the $N$ moments of the time-like 
evolution kernels and the hard-scattering coefficient functions by re-deriving them 
from their counterparts in momentum space. 
We find full agreement with the results given in the literature. 
The results obtained with our time-like evolution code are found to agree 
with the {\sc Mela} package after correcting some obvious inconsistency in generating 
their benchmark numbers.

On the phenomenological side, we have extracted new sets of parton-to-pion fragmentation functions
from a fit to electron-positron annihilation data up to next-to-next-to-leading order accuracy. 
We have compared our results to existing next-to-leading order
fits in the literature. The flavor singlet fragmentation function, which is known to be
constrained best by data, comes out very similar as in all previous fits
in both our next-to-leading and next-to-next-to-leading order analyses
whereas some small ambiguities remain for the fully flavor-decomposed fragmentation functions.
While the quality of our fits to electron-positron annihilation data
was already acceptable at leading order accuracy, it gradually improved upon including higher order corrections. 
In particular, the description of data at small momentum fractions $z$ at the lowest energies $Q$ is significantly better 
at next-to-next-to-leading order accuracy. 
In addition, leading order fits are found to explore regions in the parameter space which are
at the border of becoming unphysical in order to achieve the best possible fit to data.
As for the analysis of parton distributions, we expect
global fits of fragmentation functions at
next-to-next-to-leading order accuracy to become the new standard soon.

In the last part of the paper we have illustrated some salient features of the
next-to-next-to-leading order corrections to the evolution of fragmentation functions
and hadron production in electron-positron annihilation. The most important new asset is the found
reduction of theoretical uncertainties related to the choice of the factorization scale
by about a factor of two as compared to the next-to-leading order level. 
The uncertainties now match the precision of the data in most of the kinematic regime relevant
for an analysis of fragmentation functions. A similar reduction by a factor of two
was found for the size of the genuine higher order corrections
relative to calculations performed one order lower in the perturbative series, i.e.,
in the $K$-factor. The latter and the scale ambiguity tend to increase both for very large and
small values of $z$, indicating the presence and numerical relevance of large logarithmic corrections 
in the perturbative series, which eventually should be resummed to all orders.

There are several avenues one can follow to further improve the theoretical framework for the analysis
of fragmentation functions and the phenomenology of single-inclusive hadron production in
electron-positron annihilation. First and foremost, one can include the mentioned all-order resummations,
for which our code in Mellin moment space is particularly suited.
This will allow one to not only extend the range in $z$ where fits to fragmentation functions
can be performed reliably but it would also give access to other 
experimentally relevant quantities such as integrated hadron multiplicities.

As is well known and utilized in global QCD analyses of fragmentation functions at
next-to-leading order already, other processes such as semi-inclusive deep-inelastic scattering
or inclusive hadron production in hadron-hadron collisions provide invaluable information
on the flavor decomposition and the gluon fragmentation function. Full
next-to-next-to-leading order expressions for these processes are unfortunately 
not yet available but one can resort to results obtained with resummation
techniques that contain the dominant higher order terms. Again, these expression can
be most conveniently implemented numerically in terms of Mellin moments.

Finally, the treatment of heavy quark to light meson fragmentation functions in global
analyses certainly leaves room for improvement. For instance, matching conditions for a
variable flavor number scheme are only know up to next-to-leading order accuracy so far.
We plan to provide quantitative studies along all these directions in the near future.

\section*{Acknowledgments}
%
We are grateful to R.\ Sassot, W.\ Vogelsang, and A.\ Vogt
for helpful discussions and comments. We greatly appreciate extensive discussion with the authors of the  {\sc Mela}
code,  V. Bertone, S. Carrazza, and E.R. Nocera.
D.P.A.\ acknowledges partial support from the 
Fondazione Cassa Rurale di Trento.
This work was supported in part by 
the Bundesministerium f\"{u}r Bildung und Forschung (BMBF) under 
grant no.\ 05P12VTCTG and by
the Institutional Strategy of the University of T\"{u}bingen (DFG, ZUK 63). This research is supported by the US Department of Energy, Office of
Science under Contract No. DE-AC52-06NA25396 and by the DOE Early Career Program under Grant No. 2012LANL7033.


%
\end{document}